# Multicore architecture and cache optimization techniques for solving graph problems

Alvaro Tzul
Department of Computer Science
California State University Dominguez Hills
Carson, CA, USA
atzul1@toromail.csudh.edu

*Abstract:* **With the advent of era of Big Data and Internet of Things, there has been an exponential increase in the availability of large data sets. These data sets require in-depth analysis that provides intelligence for improvements in methods for academia and industry. Majority of the data sets are represented and available in the form of graphs. Therefore, the problem at hand is to address solving graph problems. Since the data sets are large, the time it takes to analyze the data is significant. Hence, in this paper, we explore techniques that can exploit existing multicore architecture to address the issue. Currently, most Central Processing Units have incorporated multicore design; in addition, co-processors such as Graphics Processing Units have large number of cores that can used to gain significant speedup. Therefore, in this paper techniques to exploit the advantages of multicore architecture is studied.**

## I. Introduction

There are two aspects that can be addressed using multicore architecture and cache optimization. One is the need to develop algorithms and programs that can take advantage of the multicore architecture and exploit the available hardware in both Central Processing Units (CPUs) and Graphics Processing Units (GPUs). The other is the requirement to improve the usage of the processor i.e. increasing the efficiency of the processor, so that more time is spent in doing computation rather than waiting for data transfers. This issue is addressed as the "Processor-Memory gap" problem [1]. It is evident even from a basic recursive algorithm for finding Fibonacci numbers, the computation space increases rapidly and becomes very wide. Also, the smaller problems that are generated as a part of the computation are independent. Although recursion presents some inherent opportunity for parallelism, but because of additional dependencies in specific cases, simple conversion of a sequential algorithm to its recursive version might not be straight forward.

Gene Amdahl proposed an estimate on the upper bound to the amount of parallelization that can be incorporated in an algorithm, known as Amdahl's law. It says that, in general, if a fraction α of an application can be run in parallel and the rest must run serially, the speedup is at most $1/(1 - α)$. Therefore, it is essential to identify the parts of an algorithm which are good candidates to have benefits from conversion into parallel counterparts.

Therefore, while parallelization of the algorithms makes efficient utilization of the available multiple cores, there are other techniques which can address the "Processor-Memory Gap" problem. If data can be made readily available for computation by the processors, then vital CPU cycles can be saved, which can immensely decrease the execution time of the algorithms. For example, problems that access large data structures might be able to implement novel techniques to keep the data in structures that can be easily stored in the cache for the entire time of the computation on them.

The rest of the paper is as follows. In Section 2, we discuss the architecture of multicore systems. Section 3 analyzes Matrix Multiplication, and shows how parallelization can be applied to parts of the sequential algorithm. In Section 4, we study Breadth First Search and Floyd-Warshall algorithms and identify the possible bottlenecks and approaches to address the above mentioned issues [1] [4]. Here we also look into the parallel versions of the graph algorithms and do the analysis of the same. The Section 5 of the paper deals with the study of MapReduce technique and its implementation feasibility on the multi-core architecture. Conclusion is provided in Section 6.

## II. Multicore CPU and GPU Architecture

Computers have traditionally had single core processors cast on a CPU chip. The core is comprised of a single set of registers with a corresponding Arithmetic Logic Unit (ALU). Input-output to this unit is done using the bus interface. Other than the main memory, which is accessible using the memory bus, the CPU chip also makes use of available on-chip memory locations called cache apart from the registers for storing temporary data. This memory hierarchy, consisting of the registers, different levels of cache, and the main memory determines the performance of the CPU while executing data intensive applications by reducing the latency introduced due to accessing of data from main memory or even the external disk drives. The architecture of such a device is shown in Fig. 1.

In case of a multi-core computer, the CPU chip consists of multiple sets of ALU and registers [16]. Each set of one ALU and register is defined as a core for the computer. The diagram in Fig. 2 shows a chip consisting of 4 cores.

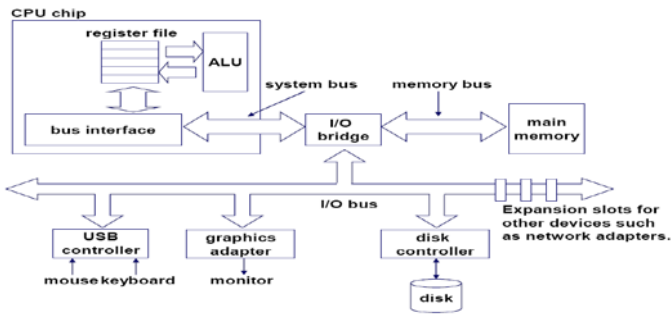

Fig. 1: CPU Architecture

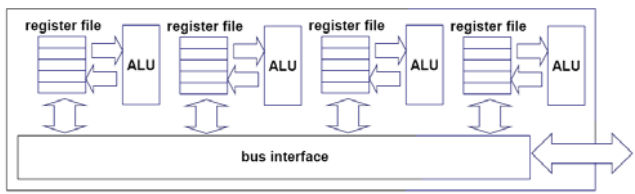

Fig. 2: Multicore chip with 4 cores

Other than multicore CPUs, the device that offers multicore architecture for computation is the GPU (Graphics Processing Unit). Computer Unified Device Architecture (CUDA), developed by NVIDIA Corporation provides platform for the graphics processing unit (GPU) to help it perform as a general purpose graphics processing unit (GPGPU) [7] [8] [22]. The GPU is referred to as the "device" and the CPU to which it is connected is called the "host".

The GPU can be controlled or accessed by programs run on the CPU and data can be transferred to the memory of the device to delegate specific tasks to be performed on it. Earlier, GPUs were specifically used to solve programs that belonged to the graphics domain. One way to utilize the computation power of the GPUs is to model general programs into equivalent graphics problem, and then solve those problems instead. But this approach is complicated and there is a lot of conversion overhead involved. Whereas CUDA allows users to directly execute programs and solve general problems in the original form. The CUDA API (Application Programming Interface) documents all the details as to how the programs that are executed on the CPU can transfer or delegate a part of the program to be executed on the GPU. It allows the programmer to define functions known as "kernels", which are executed in parallel on the device by a number of CUDA threads. There are a number of NVIDIA chips available in the market that supports CUDA. Nvidia C1060 is one of the basic chips available. It has 240 streaming processor cores in it, the frequency of the processor cores is 1.3 GHz, and there is a dedicated memory of 4GB.

CUDA provides a large number of threads that can be executed simultaneously on the cores of the device. To be able to maximize the utilization of the available hardware, it is necessary to have parallel versions of the problems that are expected to have performance gain by solving on multi-core architecture. Therefore, developing parallel algorithms for significant and basic algorithms in the areas of graph theory have been the focus [13].

The performance gain provided by CUDA lies in the fact that it can execute a large number of instructions simultaneously. But, along with this opportunity to be able to achieve significant speedup, there is the challenge of minimizing the latency introduced due to the accessing of the data elements from the memory. As in most of the CPUs of today, there is also a hierarchy of memory for the GPUs. Similar to the registers, L1, L2 and other levels of cache which are used to hide the latency introduced by the memory accesses, the GPUs have their own registers, texture cache and other forms of cache, which can be used to bring the data "closer" to the GPU for fast execution [12]. Hence, other than being able to utilize the parallel threads provided by CUDA, exploring means to be able to make use of the cache memory by using either prefetching or other techniques, like using unconventional data structures is also of primary importance in this area of research.

### III. MATRIX MULTIPLICATION

Matrix-multiplication is one of the fundamental operations in computer science. A number of problems like solving linear equations and computer graphics make extensive use of it. Therefore, if any significant improvement can be achieved in reducing the execution time of matrix multiplication, then all the other related problems would benefit from it. Below is the basic sequential algorithm for it. This algorithm multiplies the elements of matrix A with the elements of matrix B, and stores the results in the elements of a new matrix C. Here it assumes that all the matrices are square matrices i.e. they have the same number of rows and columns. All matrices can be converted to square ones by padding extra rows or columns of elements with zero values.

```
MATRIX-MULTIPLY (A, B)
1 n = A.rows
2 let C be a new n x n matrix
3 for i = 1 to n
4     for j = 1 to n
5         cij = 0
6         for k = 1 to n
7             cij = cij + aik * bkj
8 return C
```

The above algorithm can be parallelized to do some of the computations in parallel. Although it is complex to parallelize the innermost loop with various race conditions occurring, but improvements are straightforward in case of the outer loops. The parallel version of this algorithm would be as follows.

```
PARALLEL-MATRIX-MULTIPLY (A, B)
1 n = A.rows
2 let C be a new n x n matrix
3 parallel for i = 1 to n
4     parallel for j = 1 to n
5         cij = 0
6         for k = 1 to n
7             cij = cij + aik * bkj
8 return C
```

The basic computation going on can be shown using the following diagram. Here, the element c(1,1) of the resultant matrix is computed by multiplying the first row of matrix 'A' and the first column of matrix 'B'. As we can see form the figure, this computation is independent of the other elements in the given matrices. Hence, in the above parallel version of the algorithm, all the rows and columns of the matrices are split as shown below, and hence the elements of the resultant matrix can be calculated in parallel. Therefore, in this case instead of using a single processor core, $n^2$ processor cores can be used to do the above calculations.

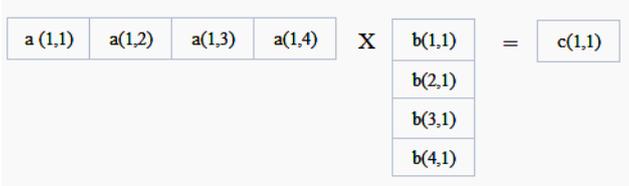

Analysis: As mentioned above, it is not straightforward to introduce parallelism inside the innermost 'for' loop of the sequential algorithm. The reason is as follows. If, after splitting of the matrices into rows and columns they are further subdivided into single elements, then the innermost loop can be calculated in parallel. But the operation in step 7 of the algorithm requires a cumulative addition. So, if different cores or threads perform the multiplication and update the value of ci,j simultaneously, then there can be errors due to the presence of race conditions. Although this situation can be avoided by synchronizing the threads or cores, but then the basic advantage of doing this part of the computation in parallel is nullified. Hence, the innermost loop is not modified using the above technique.

But, one other concern in solving the matrix multiplication problem is the size of the data. As we can see, for the simple algorithm, the requirement for the data to be in the main memory is quite stringent. In the memory the elements of the matrix are stored in row major order. So, accessing the rows of a matrix at a time is simple in that respect. Also, the chance that the data elements of a single row are present in the cache is also high because of the contiguous location of the elements in the memory. But, while accessing the data of the matrix column wise, the elements of the matrix are not contiguous and might span across many pages in the memory, and the probability of all the elements being present in the cache is low. Also, if the data size is larger than the main memory, then there would be a lot of data transfers between the external storage and the main memory. Another level of data transfer occurs between the main memory and the cache. If the whole of the data on which the recent computation is being done is in the cache, then there is significantly less number of CPU cycles spent in doing the data transfers which are used in doing valuable computation instead.

Memory is organized in the computer in many levels of hierarchy. The fastest and closest to the processor and located on-chip are the registers and L1 cache. There are different levels for the cache. L1, L2 and nowadays even L3 cache levels are available in the computer systems. The per byte storage cost of the memory located closer to the processor is much more than the ones located further away. The following table compares the amount of CPU cycles spent in accessing data from the various levels of the memory hierarchy.

| Memory Unit | Hierarchy Level | Typical Access Time |
|---|---|---|
| Registers - | Level 0 | inside CPU, less than one CPU cycle |
| L1 cache - | Level 1 | on CPU die, 1-2 CPU cycles |
| L2 cache - | Level 1 | on or off CPU die, 5 cycles |
| L3 cache - | Level 1 | external to CPU, 10-20 cycles |
| RAM/Main memory - | Level 2 | 100-1000 cycles |
| Hard disk | Level 3 - Outboard storage | over 1e6 cycles |

For example, considering an Intel Xeon Processor, with a clock speed of 2.53 GHz, the time for 1 CPU cycle is 0.395 ns. Therefore, for each of the above Memory units, the access time is given in the following table:

| Memory Unit | Access Time in ns |
|---|---|
| Registers - | < 0.395 |
| L1 cache - | 0.395 - 0.79 |
| L2 cache - | 1.975 |
| L3 cache - | 3.95 - 7.9 |
| RAM/Main memory - | 39.5 - 395 |
| Hard disk | $0.395 \times 10^6$ |

Therefore, to make efficient use of the cache, it is imperative that the data is as close to the processor as possible. Hence, we can even use simple modifications to achieve significant efficiency gains. Just by dividing the matrices into smaller parts and then multiplying those using different cores can be such an approach. The smaller size of the now divided data would help to keep it in the cache, whereas multiple cores can decrease the execution time by working in parallel.

For example, let C be the resultant matrix from the multiplication of the A and B matrices.
$$C = AB \qquad A, B, C \in R^{(2^n \times 2^n)}$$
If the matrices A, B are not of type $2^n \times 2^n$ then we can fill the missing rows and columns with zeros. We partition A, B and C into equally sized block matrices.

$$A = \begin{matrix} A_{1,1} & A_{1,2} \\ A_{2,1} & A_{2,2} \end{matrix} \qquad B = \begin{matrix} B_{1,1} & B_{1,2} \\ B_{2,1} & B_{2,2} \end{matrix}$$
$$C = \begin{matrix} C_{1,1} & C_{1,2} \\ C_{2,1} & C_{2,2} \end{matrix}$$

with $A_{i,j}, B_{i,j}, C_{i,j} \in R^{(2^{(n-1)} \times 2^{(n-1)})}$

then, $C_{1,1} = A_{1,1}B_{1,1} + A_{1,2}B_{2,1}$
$C_{1,2} = A_{1,1}B_{1,2} + A_{1,2}B_{2,2}$
$C_{2,1} = A_{2,1}B_{1,1} + A_{2,2}B_{2,1}$
$C_{2,2} = A_{2,1}B_{1,2} + A_{2,2}B_{2,2}$

So, here the results of $C_{1,1}$, $C_{1,2}$, $C_{2,1}$ and $C_{2,2}$ can be calculated using different cores in parallel. In this case, as the data is divided into smaller matrices, the probability of the data being in the cache at the time of computation is much more than the previous case. Of course there can be added improvements by using recursion to subdivide the multiplications further and engaging more available cores and achieve better parallelism [26].

Strassen's Method: The above approach concentrates on the availability of multiple processors to do the job simultaneously. There have been many other improvements made to the sequential algorithm, and one of the significant ones is Strassen's method. With multicore architecture similar improvements can also be applied to the Strassen's method. The complexity of the original matrix multiplication algorithm is $O(n^3)$. This is because of the number of multiplications used as the basic operations. Using modifications to the algorithm, Strassen's method reduces the number of multiplications from 8 to 7, and thereby the complexity becomes $O(n^{2.81})$.

$$\begin{bmatrix} C_{11} & C_{12} \\ C_{21} & C_{22} \end{bmatrix} = \begin{bmatrix} A_{11} & A_{12} \\ A_{21} & A_{22} \end{bmatrix} \times \begin{bmatrix} B_{11} & B_{12} \\ B_{21} & B_{22} \end{bmatrix}$$

As in the above case, the matrices can be divided into 4 equal parts. Now, to calculate each part of the C matrix, there are 2 multiplications required. So, there are total 8 multiplications required in this case. Strassen's method modifies this calculation, and reduces it to 7 multiplications, though the number of additions increases. But, as the computation cost of the addition is much less than that of a multiplication, so it provides an improvement over the previous scheme.

The following equations show how the calculation is done.

$$P_1 = (A_{11} + A_{22})(B_{11} + B_{22})$$
$$P_2 = (A_{21} + A_{22})B_{11}$$
$$P_3 = A_{11}(B_{12} - B_{22})$$
$$P_4 = A_{22}(B_{21} - B_{11})$$
$$P_5 = (A_{11} + A_{12})B_{22}$$
$$P_6 = (A_{21} - A_{11})(B_{11} + B_{12})$$
$$P_7 = (A_{12} - A_{22})(B_{21} + B_{22})$$
$$C_{11} = P_1 + P_4 - P_5 + P_7$$
$$C_{12} = P_3 + P_5$$
$$C_{21} = P_2 + P_4$$
$$C_{22} = P_1 + P_3 - P_2 + P_6$$

Although, apparently it might seem as if this method performs better, but actually in case of multicores the case is opposite. Here, as it can be seen, to reduce the number of multiplications, there have been a number of additions introduced. Also, it can be noted, that the calculation of the parts of the final matrix can only be done after the initial parts are completed. So, if this is implemented in case of multicores, then the cores designated to compute the final additions, must be synchronized with the cores doing the multiplications and calculation of the P's. But, this is an added overhead, and also the extra additions don't come for free. So, this ultimately reduces the usage from simultaneously 8 cores to 4 cores. Hence, although theoretically Strassen's method is an improvement on the general approach, but the former is a better candidate to exploit the resources in a multicore environment [28].

## IV. GRAPH ALGORITHMS

Graph algorithms have a lot of real life applications based on them [17] [20]. Starting from network analysis, to data mining and design of electronic circuits, all make use of these algorithms [18]. But some of these problems deal with large data sets, and computing on them becomes an issue [14] [15]. Therefore, if the data can be divided into smaller problem sets, and computed, then the tasks become easier to solve. Compressing the graph to reduce the size is also an option [9] [10] [21]. But the other concern with graph algorithms is their random data access, whereby simple techniques like tiling and blocking which work in the case of other algorithms, is not always a good choice to address the matter here [19] [23]. Therefore, modifying the algorithms to make better use of the cache is an important area of research interest.

Breadth First Search (BFS) is one of the fundamental graph algorithms which have many applications. Although it is argued that BFS is inherently sequential, but using modifications to the original algorithm, some parts of the code can be executed in parallel. The BFS problem can be stated as the one to find the minimum number of edges needed to be traversed to reach from the source vertex to every other vertex in a graph. The basic steps of the algorithm are given below.

**Breadth First Search** (Graph G(V,E), Source Vertex S)
1. Add root node S to the Queue.
2. Remove a node from the Queue and examine it.
    a. If required element found in this node, Return "Node id". Goto Step 5.
    b. Else add to Queue any successors that have not yet been discovered.
3. If Queue is empty, every node on the graph has been examined – Return "not found".
   Goto Step 5.
4. Goto Step 2.
5. Exit.

Analysis: There is an important concept of "inherently sequential" algorithms. These algorithms are members of the complexity class P-Complete. Also, there is another set of algorithms that belong to the class NC which can be easily parallelized. Now, generally the BFS is implemented using the queue data structure. If following the same algorithm, the data structure is changed to a stack, then it becomes a Depth First Search algorithm. But, there are versions of the BFS which can make use of the stack in doing the graph traversal. According to studies already done, the version of the BFS which makes use of the stack is inherently sequential. So, no matter how many processors are at the disposal of the code, it cannot make use of more than one processor, and no speedup can be achieved in a multicore environment. But, on the other hand, the BFS that makes use of the queue, belongs to the NC class, and can be easily parallelized. Therefore, there exists algorithms that make certain changes to the original version, and the result is a parallel version of the same that can take full advantage of the multicore architecture.

Parallel BFS: The BFS can be converted to a parallel version by modifying the above algorithm. The basic idea behind it is, expanding and updating the child nodes of the node being processed in parallel. So, the modified algorithm would be as follows:

**P-Breadth First Search** (Graph G(V,E), Source Vertex S)
1. Add root node S to the Queue.
2. Do in parallel for all the nodes in the Queue
    a. Remove each node from the Queue and examine it.
        1. If required element found in this node, Return "Node id". Goto Step 6.
        2. Else add to Queue any successors that have not yet been discovered.
3. Synchronize.
4. If Queue is empty, every node on the graph has been examined – Return "not found".
   Goto Step 6.
5. Goto Step 2.
6. Exit.

To avoid the Breadth First Search converting into a Depth First Search, there is a need for synchronization after each level, and this is done in Step 3 of the modified algorithm. If this is not done, then the thread or the processor updating the node with the least number of neighbors would finish first, and then go on to process and update the next level in the graph, while the other threads or processors are still working on the previous level. This synchronization is an added overhead in this algorithm. Therefore, as with the case of all tradeoffs, if the problem size is small, then the overhead might contribute more work to be done than the original problem itself.

Issues with Parallel BFS: It is important to analyze the effective parallelizability of the above algorithm. In step 2, all the neighbors of a node are updated in parallel using threads or multicores. Now, in case of dense graphs, more than one node might have the same neighboring node. So, in that case it might result in race condition while updating the weights. This can be avoided by using semaphores, but then the resulting solution is not absolutely parallel. Though this is a bottleneck, but in cases where all the neighbors of the nodes already in queue are distinct, the above algorithm indeed executes in parallel.

**Floyd-Warshall Algorithm:**
The Breadth First Search technique helps to find the distance of all the vertices in a graph from the source vertex. This algorithm can be referred to as the "single source shortest path" algorithm. Another important type of algorithm that has many practical applications is the "all-pair shortest path" algorithm, and Floyd-Warshall is one such algorithm. After executing it on weighted, directed graphs, the distance between all the pairs of vertices can be found out. This algorithm belongs to the "dynamic programming" class of algorithms. Here at each step, values are updated based on those already calculated in previous iterations. The iterative version of the algorithm is given as follows:

Floyd-Warshall (W)
   // Let N be the problem size of W
1. $D^0 \leftarrow W$
2. for $k \leftarrow 1$ to N
3.   for $i \leftarrow 1$ to N
4.     for $j \leftarrow 1$ to N
5.       $d^k_{ij} \leftarrow \min( d^{k-1}_{ij}, d^{k-1}_{ik} + d^{k-1}_{kj} )$
6. return $D^N$

It can be observed that an entire NxN array is being updated after each step of the outer loop. For most practical purposes, the size of the array would be large enough to not fit in the cache entirely. This would lead to a huge overhead in terms of data movement between the memory and the cache for the required calculation. Now, just as in the case of Matrix Multiplication, executing some of the steps in this algorithm in parallel to achieve better efficiency can be thought as an option. The basic idea here is to divide the FW Iterative algorithm into a FW Recursive algorithm. But, due to additional dependencies among the data in the algorithm such a transformation from an iterative to a recursive algorithm is not straight forward. In [1] the authors propose an algorithm, which is recursive, and prove that the results of their algorithm and the original one are the same. The algorithm is as follows:

Floyd-Warshall Recursive (A, B, C)
1. if(base case)
2.   FWI (A, B, C)
3. else
4.   FWR ($A_{11}$, $B_{11}$, $C_{11}$);
5.   FWR ($A_{12}$, $B_{11}$, $C_{12}$);
6.   FWR ($A_{21}$, $B_{21}$, $C_{11}$);
7.   FWR ($A_{22}$, $B_{21}$, $C_{12}$);
8.   FWR ($A_{22}$, $B_{22}$, $C_{22}$);

9.     FWR ($A_{21}$, $B_{22}$, $C_{21}$);
10.    FWR ($A_{12}$, $B_{12}$, $C_{22}$);
11.    FWR ($A_{11}$, $B_{12}$, $C_{21}$);

Analysis: The basic idea of the above algorithm is to divide the original problem into smaller ones to take advantage of the data being present in the cache to avoid wasting CPU cycles in getting the data from the memory. But the main concern here is the extent to which the above modified algorithm achieves parallelism. The algorithm has additional dependencies to take care of. So, the ordering of the recursive calls in the above algorithm is essential for the correct execution of the same. Now, the arguments A, B and C to the algorithm refer to the same matrix and the subparts like A11, B11 and C11 refer to the same elements. So, in the first call FWR (A11, B11, C11), the part A11 is updated and in the next call FWR (A12, B11, C12), B11 refers to the updated A11 from the first call. In this way, all the calls are dependent on the previous calls to start the processing [27].

So, the point here is, though the algorithm performs better by just dividing the problem into sizes that fit in the cache, but the recursive parts cannot be executed in parallel on multicores if they are available. For example, the computations going on here are different from the matrix multiplication algorithm discussed earlier in this report. In case of the matrix multiplication all the calculations are independent, the resultant parts of the matrix do not depend on the other parts of the result, but are just calculated based on the input values. But here the case is different as each recursion modifies the input to the successive recursions. Hence, this algorithm uses the cache to its advantage, but does not utilize multicores in its implementation.

In this paper [1], the implementation does not divide the data till the number of elements become one as in most other recursive algorithms. Here, the base condition is reached whenever the current data fits in the cache. So, it is normally some multiple of the number of elements that fit in a cache line. As a result of this scheme a speed-up of 2 is achieved on the machines the implementations are carried out. Also in this paper [1], the authors take into account the data layout to avoid conflict misses in the cache. Other papers avoid taking into account this issue just by making an assumption that the cache is fully associative.

Cache-aware and Cache-oblivious Algorithms: There are two types of cache optimization algorithms – Cache-aware and Cache-oblivious. In the first case, the size of the cache is known to the program, so it tries to modify the data access patterns in a way which optimizes the usage of the cache, or in other words, minimizes cache misses, minimizes cache pollution and therefore reduces the data access time. In case of Cache oblivious algorithms, the size of the cache is not known. The algorithm keeps dividing the data into smaller parts, and thereby ensures that when a step of the algorithm is solved, the data it refers to is already in the cache.

But, there is an important concern with the cache-oblivious algorithms. If the data is divided till it becomes a single element, then the overhead to manage the divided data and to combine it back preserving all the dependencies of the original problem can severely degrade the performance of the problem.

So, combining the basic ideas of both the algorithms, a hybrid method can be thought of. In this case, the algorithm would be pseudo cache aware. The problem would be divided into parts, and it would stop when it hits the base case, and in this situation, the base case would be a "general" cache size. This "general" size can be calculated by taking into account the different available architectures.

Prefetching: Other than using cache optimization techniques to improve the performance of the cores, there is another mechanism which helps in effective utilization of CPU cycles – prefetching. Prefetching is basically the method which makes data available before it is actually required for computation by the CPU by bringing it to the various levels of the cache from the memory, thereby saving CPU cycles by avoiding to wait for the data to be fetched from the main memory in the middle of a computation. This can be a significant gain as seen from the information available in the earlier table, where it shows the time to access data from the main memory is about 20 times slower compared to getting it from the cache.

Prefetching can be of two types – hardware prefetching and software prefetching. The name hardware prefetching is used when the CPU does the prefetching on its own without any instruction from the programmer. This is basically a system prefetch, and the effects of this are profound when the data access pattern is regular, like bringing in data from an array. But in cases like graph algorithms, where the data access pattern is random, this can actually degrade the performance.

The other type of prefetching is the software prefetching. Here, the programmer puts in explicit instructions within the code to do the prefetching. In case of programs, where the programmer is aware of the data structure being used, and the stride length of the data, then data prefetching might be a viable option. But the issue with software prefetching is cache pollution. If a data that is being currently computed upon is evicted from the cache to bring in the data that will be used in the future, then the penalty for such an incident can lead to an overall degradation of the performance of the CPU.

There are options which can be used to enable or disable the hardware prefetching. In [4], the authors show some interesting results. They choose algorithms with cache optimization techniques like the cache aware algorithms and their normal counterparts. With the hardware prefetching option disabled, the execution time of the cache aware algorithms is better than the normal ones. But, when the hardware prefetching is enabled, the results are not as expected. The execution time of the normal algorithm is less than the cache aware ones. Here, increased complexity in the data access patterns of the cache aware algorithms nullify the effectiveness of the hardware prefetcher, whereas the original algorithms can make use of the hardware prefetching capability and outperform their improvised counterpart. Therefore, this is an important consideration to be kept in mind while trying to improve algorithms. Care should be taken to take advantage of the hardware prefetcher, and complicated data access should be avoided.

## V. MAPREDUCE

MapReduce is a technique employed for analyzing large amounts of data in a reasonable amount of time using a distributed approach [3]. The model consists of two sets of functions – Map and Reduce. Both of these functions are written by the users. The Map function takes the user input, does the initial computation, and forwards the data to the reduce function. The reduce functions does its part of the calculation and writes the output to an external file. In some cases, to get the required output more iteration is needed. So the output of the reduce functions in these cases are again returned back as input to the Map functions. The proposed interface is mainly designed for implementation in a clustered environment. But, the same technique can be applied to other architectures [11], specifically multicore systems - the one we are interested in.

The MapReduce library contains other functions that make the computations easily parallelizable. Through the interface, the user can provide values to manage the amount of resources used, and thereby have a direct control on the amount of parallel computation desired. The basic computations are simple, and the real complexity comes from the large volume of data being processed. At large business organizations, it is required to process a large number of data on a daily basis, like the data gathered from the web crawlers. Therefore, partitioning the data, fault tolerance, minimizing data movement between the machines and optimizing the number of machines required are the priorities of this model.

The model of MapReduce involves a number of workers and a master, each of which is a node in the cluster environment. MapReduce also requires synchronization among the workers. One group of workers can start processing data that is produced by another group. But this operation can be successfully executed only if there is proper communication among the different set of workers. The situation where a group tries to process data that has not been completely produced by the other group can lead to unexpected results. This is where the master plays an important role. The master communicates with both the groups and sends timely signals to the groups as and when required and also checks the proper functioning of the workers by ping and response mechanism.

Analysis: The implementation of the MapReduce model takes into consideration many nuances other than parallelizability.

a) Fault tolerance: In case of implementation of MapReduce in the clustered environment, an important consideration is checking for the failures of the workers to incorporate fault tolerance. The master pings the workers at regular intervals, and if there is no response consecutively for a specific number of times, then the master assumes that the worker node has failed. It then takes care of reassigning the task to another available worker. The failure of the master, which is a single node, is quite rare. To take care of this situation the master has a back-up of the completed work that is updated at a specific interval. If the master fails, then another node which then becomes the master can proceed from the last check point as indicated on the backup.

b) Partitioning function: Generally, the data set being processed is quite large, and there are a number of Map functions processing different parts of the data in parallel. Therefore, partitioning the data to be distributed among the processors is an important issue. It can be either a simple partition by just dividing the data equally among the number of processors available, or divide it into more parts, and assign more than one part to be processed on each processor. Although the former scheme might have less overhead, but the later provides better fault tolerance. If one of the nodes fails before completing the work, then it has to be redone. In the second case the work needed to be done again would be only that part which failed, whereas in the first case, the amount of work done again would be much more.

c) Local execution: Network bandwidth is expensive, so minimizing data movement between the different nodes in the cluster environment is a priority. In the cluster environment, data is duplicated and kept in more than one node. The master tries to schedule the job of processing the data on a node which already has a copy of the data, or else allocate it to the nearest available node. Also, if data is computed on a machine, and the results need to be debugged, then it is always better to do the debugging on the same machine that calculated the data in the first place, because chances are high that the computation would benefit from the data that might still be in the cache of the same machine.

MapReduce and its feasibility on multi-core architecture: The implementation of MapReduce in the multicore environment has a significant correspondence with its cluster counterpart. In the context of multicores, the workers can be thought of as the cores in the GPU or the different cores available in a multi-core CPU machine. The different threads in the GPU i.e. the device can act as the workers, and the main CPU i.e. the host can act as the master. The initial data can be transferred to the memory of the GPU and both the intermediate and the final output can be written in there. Finally, the output files can be transferred back to the memory of the CPU.

Fault tolerance can be an issue in case of GPUs. In this case, as the threads are functioning as the workers, then it is important to study what are the chances of threads failing in a multicore system's GPU. Also, while considering a multicore system, if we are in the domain of computers with multiple CPUs and no GPUs, then the probability of failure of a CPU core and its analysis is important. The case where the master fails, although has a very low probability, is also covered in the machine with GPU, where the failure of the master is indicated by the failure of the machine as a whole. But, in case of a multi CPU machine, the failure of the master might not be that obvious.

Local execution can be thought of having similarities to the idea of processor affinity. In case of multicore CPUs, the master scheduling the jobs among the workers can take note of which jobs are being processed by which CPU core. In case of debugging, the same core should be preferred, as there are chances that the cache memory has some of the data in them already to do much faster computation compared to a new core being chosen, and the delay associated with reading from the memory is involved.

Therefore, implementing the MapReduce in a multicore environment, especially with multicore GPUs is feasible as evident from the discussion above.

Matrix Multiplication using MapReduce: The basic model of MapReduce incorporates parallelism by dividing the computation among available resources. This architecture can be used to achieve the effect of multithreaded matrix multiplication. The fundamental idea here is to modify the Map and Reduce functions to divide the matrix data and do calculations in a distributed manner. The data store, Map and Reduce functions would be modified as follows:

Data Store: This would contain the input matrices that need to be multiplied and also the resultant output matrix. The Map functions would be able to access it using local access or remote access as the case might be. Also, if the size of the input data is large, more than one data store might be required.

Map: This would access the data store(s), and rewrite the data into the intermediate store rearranging it according to the row and column numbers. There would be two types of Map functions. The output of the first type of Map function would be like (Row #, {Column #, data element}), and the output of the second type would be (Column#, {Row #, data element}). So, it basically divides the input matrices into rows and columns.

Reduce: There would be $2^n$ Reduce functions, where n would depend on the amount of parallelization required. If the input matrices are divided into smaller matrices, then n would be large, and it would employ more number of cores or processors. In the base case, n should be 2, so all the matrices are divided into 4 equal parts and the rest of the computation can be carried out as discussed in Section 3 of this report. This function would compute the multiplication of the row elements with the corresponding column element, and write the value in the resultant matrix.

Hence, by properly writing Map and Reduce functions, the MapReduce model can be used to compute matrix multiplication in parallel.

## VI. CONCLUSION

In this paper we look at some of the basic problems in fundamental areas of computer science to exploit the available multicore architecture. We study and discuss matrix multiplication and the different ways to execute it in a parallel environment. We analyze the various approaches that can be used to obtain the desired result. Using similar methodology, we try to identify the sections of graph algorithms that can be executed in parallel. Here we study and discuss the Breadth First Search and Floyd-Warshall algorithms. Apart from the improvements achieved by shifting from a sequential model to a parallel one, we also consider the aspect of added efficiency by making improvements in the data access patterns by modifying the data structures and using other schemes like prefetching to better use the cache. We also look into the feasibility of executing MapReduce in a multicore environment with the advantages and challenges associated with it. Finally, we analyze and discuss the implementation of matrix multiplication using the MapReduce technique.